\documentclass[%
 reprint,
 superscriptaddress,
 amsmath,amssymb,
 aps,
prb,
]{revtex4-2}

  

\makeatletter
\def\@hangfrom@section#1#2#3{\@hangfrom{#1#2}#3}
\def\@hangfroms@section#1#2{#1#2}
\makeatother

\usepackage[dvipsnames]{xcolor}
\usepackage{physics}
\usepackage{graphicx}
\usepackage{dcolumn}
\usepackage{bm}

\usepackage{float}
\usepackage{booktabs}
\usepackage{multirow}
\usepackage{placeins}

\usepackage{xspace}

\usepackage[figuresleft]{rotating} 

\usepackage[version=3]{mhchem} 
\usepackage{caption}
\usepackage{subcaption}
\usepackage{chemformula}
\usepackage[separate-uncertainty = true]{siunitx}
\DeclareSIUnit\atom{atom}
\DeclareSIUnit\angstrom{\text{\AA}}
\usepackage{nameref}    
\usepackage{multirow}   

\usepackage{titlesec}

\titleformat{\subsection}
  {\normalfont\slshape\filcenter}  
  {\thesubsection}{1em}{}  

\begin{document}

\preprint{Draft}

\title{Optimizing Machine Learning Potentials for Hydroxide Transport: Surprising Efficiency of Single-Concentration Training}

\author{Jonas Hänseroth}
\affiliation{Theoretical Solid State Physics, Institute of Physics, Technische Universität Ilmenau, 98693 Ilmenau, Germany}

\author{Christian Dreßler}
\email{christian.dressler@tu-ilmenau.de}
\affiliation{Theoretical Solid State Physics, Institute of Physics, Technische Universität Ilmenau, 98693 Ilmenau, Germany}

\date{\today}

\begin{abstract} 
\noindent \textbf{Abstract} We investigate the transferability of machine learning interatomic potentials across concentration variations in chemically similar systems, using aqueous potassium hydroxide solutions as a case study. Despite containing identical chemical species (\ce{K+}, \ce{OH-}, \ce{H2O}) across all concentrations, models fine-tuned on specific KOH concentrations exhibit surprisingly poor transferability to others, with force prediction errors increasing dramatically from $~$\SI{30}{\milli\eV\per\angstrom} (at training concentration) to $~$ \SI{90}{\milli\eV\per\angstrom} (at very different concentrations). This reveals a critical limitation when applying such models beyond their training domain, even within chemically homogeneous systems. We demonstrate that strategic selection of training data can substantially overcome these limitations without requiring extensive computational resources. Models fine-tuned on intermediate concentrations (\SI{6.26}{\mole\per\liter}) exhibit remarkable transferability across the entire concentration spectrum (0.56-\SI{17.89}{\mole\per\liter}), often outperforming more computationally expensive models trained on multiple concentration datasets. This approach enables accurate simulation of hydroxide transport dynamics across varying electrolyte conditions while maintaining near-quantum accuracy. Our simulations further reveal the emergence of hydroxide-hydroxide hydrogen bonding at high concentrations - a phenomenon not explicitly represented in dilute training data but successfully captured by our intermediate-concentration model. 
This work establishes practical guidelines for developing broadly applicable machine learning force fields with optimal transferability, challenging the assumption that diverse training datasets are always necessary for robust performance in similar chemical environments.

\end{abstract}

\maketitle

\section*{Introduction}
Hydrogen production using renewable energy at scale is essential for transitioning to sustainable industrial processes and energy generation. Among available methods, water electrolysis has emerged as a leading approach, with anion-exchange membrane (AEM) technology offering particular advantages. AEM water electrolysis combines high efficiency with the ability to use abundant, low-cost electrode materials like iron and nickel.\cite{henkensmeier2020, hren2021, park2019, leng2012, dekel2018} This contrasts with proton exchange membrane (PEM) electrolysis, which operates in acidic environments, requires PFAS-free materials, and depends on scarce, expensive noble metal catalysts such as platinum and iridium.\cite{alia2021, zou2015, schalenbach2018}

A significant challenge for AEM electrolysis is the relatively poor hydroxide ion conductivity in membrane materials.\cite{wijaya2024, henkensmeier2020} To address this limitation, computational simulations of the liquid electrolyte in the system, aqueous potassium hydroxide solution, are required to provide valuable insights by predicting hydroxide mobility.\cite{han2014, wang2018, dekel2018_sim, karibayev2022} These simulations lay the groundwork for optimizing of AEM materials at both the design and synthesis stages, thus accelerating the development of higher-performing membranes. 

While force field molecular dynamics (FFMD) simulations offer computational efficiency and have been extensively validated for water and hydroxide ions, they cannot adequately model hydroxide mobility.\cite{delucas2024} This limitation stems from the fact that hydroxide transport involves complex bond-breaking and bond-forming processes that only quantum chemical methods can accurately capture.\cite{zelovich2024, tuckerman2006acs, ouma2022} To overcome these computational barriers, researchers have developed two alternative approaches. The first integrates MD simulations with complementary techniques such as Monte Carlo simulations, extending the accessible timescales of ion dynamics from nanoseconds to milliseconds.\cite{dutta2024, dressler2016, kabbe2016, kabbe2017, haenseroth2025ohlmc} The second approach employs machine-learning-based interatomic potentials (MLIP), enabling hydroxide ion mobility simulations at timescales comparable to classical MD while maintaining near \textit{ab initio} accuracy.\cite{hellstrom2018, karibayev22_aemmlff, jinnouchi23_proton} However, this method remains relatively unexplored for AEM materials and aqueous potassium hydroxide solutions.

In this article, we focus on the second approach. Predicting atomic forces with machine-learned force fields at \textit{ab initio} accuracy represents a central goal of various machine learning methods. Gaussian Approximation Potentials have already demonstrated success in molecular dynamics simulations.\cite{bartok2010, unke2021, friederich2021, grunert2025, reiser2022} Significant advances have occurred through the evolution from traditional descriptors like atom-centered symmetry functions and smooth overlap of atomic positions descriptor to the atomic cluster expansion (ACE), which provides more flexible and efficient representation of atomic environments.\cite{behler2007, bartok2010, drautz2019} Additionally, equivariant graph neural networks have further enhanced these models' predictive capabilities.\cite{reiser2022,batzner2022} 

Another significant advancement is the development of machine-learned force fields (MLFFs) trained on extensive datasets, creating foundation models. These models offer broad applicability across diverse materials while allowing fine-tuning for specific systems.\cite{batatia2023, yang2024} This approach has produced highly accurate and versatile foundation models that substantially expand the potential of MLFFs in molecular simulations.\cite{batatia2022, batatia2023, kovacs2023}

Our research focuses on MACE, one of the most promising publicly available machine-learning software packages for predicting many-body atomic interactions and generating force fields.\cite{batatia2022, batatia2023} We investigate how fine-tuned MACE models transfer when simulating potassium hydroxide (KOH) in aqueous solution across varying concentrations and temperatures.

Both \ce{OH-} and \ce{H3O+} exhibit enhanced mobility compared to water molecules due to the Grotthuss mechanism. This process consists of two key steps: a proton transfer, where a proton hops along hydrogen-bonded particles, followed by a reorientation step that reorganizes the surrounding water structure to enable subsequent transfers.\cite{grotthuss, marx2006, tuckerman2006acs, tuckerman1995jpc} Through this mechanism, charge transport occurs rapidly without requiring significant molecular diffusion.\cite{tuckerman2010rev}

A critical challenge in anion-exchange membrane electrolysis is understanding hydroxide ion transport under realistic operating conditions. Electrolyte concentration and operating temperature directly influence ionic conductivity, reaction kinetics, and membrane stability in electrolyzers.\cite{henkensmeier2020,franck1965} Therefore, understanding trends of  hydroxide ion mobility within the standard KOH concentration range of 3–10\,wt.\si{\percent}, as well as at higher concentrations up to 30\,wt.\si{\percent}, and at temperatures between 50 and \SI{70}{\degreeCelsius}, is essential for optimizing AEM performance.\cite{henkensmeier2020, miller2020, busacca22019}

Machine-learning-based force field simulations provide a powerful tool for capturing the complex interactions between hydroxide solvation and hydrogen bonding dynamics across these conditions. These simulations enable molecular-level insights into how concentration and temperature variations affect hydroxide ion transport. Understanding these relationships is crucial for identifying optimal electrolyte conditions that maximize ionic conductivity in AEM materials.

This study aims to deepen our understanding of hydroxide ion transport over extended time scales by leveraging machine learning-based force field simulations. By systematically examining how electrolyte concentration and temperature influence transport properties, we bridge the gap between theoretical models and practical operating conditions. Our work will contribute to developing broadly applicable machine learning force fields capable of simulating AEM materials at greater time and length scales, ultimately supporting the development of more efficient and durable electrolyzers for green hydrogen production.

\FloatBarrier
\section*{Method}
The Python package MACE provides an accessible framework for obtaining fine-tuned foundation models for molecular simulations.\cite{batatia2022,batatia2023} In this study, we employed the MACE-MP-0 foundation model, which was pre-trained on the comprehensive Materials Project database, giving it broad capabilities for simulating various chemical systems.\cite{batatia2023,jain2013,ong2015}

\begin{figure*}[!ht]
		\centering
		\includegraphics[width=\textwidth]{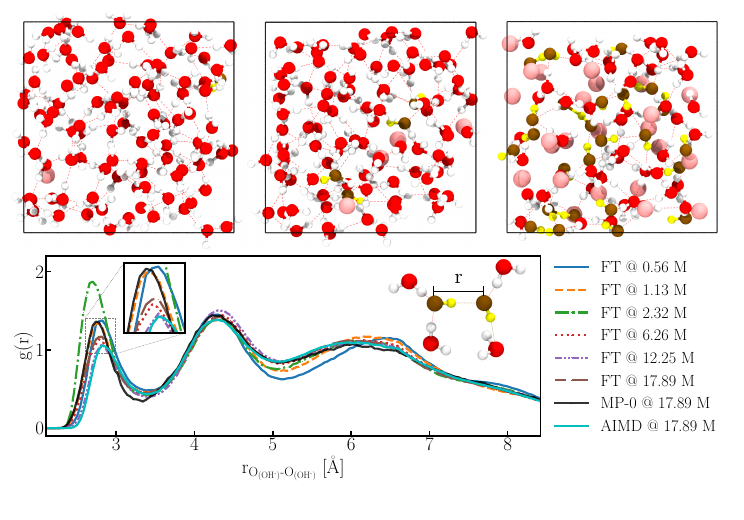}
        \caption{\textbf{Upper panel}: Snapshots from the AIMD trajectory of the aqueous potassium hydroxide solution at different KOH concentrations at \SI{333}{\kelvin}: \SI{0.56}{\mole\per\liter} (left), \SI{2.32}{\mole\per\liter} (center) and \SI{17.89}{\mole\per\liter} (right). Oxygen atoms are shown in red, hydrogen in white and potassium in rose. Hydroxide moieties are highlighted by marking their oxygen atoms in brown and their hydrogen atoms in yellow. Note the increasing number of ions as concentration increases from left to right. \textbf{Lower panel}: Radial distribution function of O\textsubscript{(\ce{OH-})}-O\textsubscript{(\ce{OH-})} distances in the system with $c$(KOH)=\SI{17.89}{\mole\per\liter}, as predicted by different single-concentration MACE-FT models, the MACE-MP-0 foundation model, and reference AIMD simulation. Including a snapshot of two hydroxides with a oxygen-oxygen distance with \SI{2.87}{\angstrom}.}
		\label{img:boxes}
\end{figure*}

To fine-tune the foundation models, we created two types of training datasets of different sizes. For our primary models, we extracted every 200th frame from \SI{200}{\pico\second} first-principles molecular dynamics trajectories at \SI{333}{\kelvin}, yielding 2000 frames per concentration. For our secondary set of models, we utilized shorter AIMD simulations of \SI{20}{\pico\second}, sampling every 200th frame to produce datasets of 200 frames each. Detailed parameters of these AIMD simulations can be found in \nameref{sec:comp_detail}, specifically in Table \ref{tab:aimd_KOH}. Figure \ref{img:boxes} illustrates the simulation boxes for aqueous potassium hydroxide solutions at three representative concentrations: \SI{0.56}{\mole\per\liter}, \SI{2.32}{\mole\per\liter}, and \SI{17.89}{\mole\per\liter}.

The fine-tuning was conducted across a range of concentrations and temperatures. For our 2000-frame models, we targeted KOH concentrations of 0.56, 1.13, 2.32, 6.26, 12.25, and \SI{17.89}{\mole\per\liter} at a fixed temperature of \SI{333}{\kelvin}. For our 200-frame models, we focused on four concentrations (0.56, 2.32, 6.26, and \SI{17.89}{\mole\per\liter}) across five temperatures (\SI{303}{\kelvin}, \SI{323}{\kelvin}, \SI{333}{\kelvin}, \SI{343}{\kelvin}, and \SI{363}{\kelvin}). These models trained on the smaller dataset are only used to obtain the force error shown in Figure \ref{img:transfer_temp}. The specific parameters used during fine-tuning are provided in Table \ref{tab:mace_ft}. To evaluate the accuracy of our fine-tuned models, we compared their predictions against quantum chemical reference data, with the resulting force and energy errors summarized in Table \ref{tab:mace_err}.

\begin{table}[!ht]
    \centering
    \caption{Root mean square error (RMSE) of force and energy of the fine-tuned MACE-MP-0 models for aqueous potassium hydroxide solution systems. Each model was evaluated against reference first-principles data obtained from AIMD simulations conducted with CP2K.}
    \label{tab:mace_err}
    \resizebox{0.48\textwidth}{!}{
    \begin{tabular}{l l l l | l l l l}
        \toprule
        T & $c$(KOH) & RMSE F & RMSE E  & T & $c$(KOH) & RMSE F & RMSE E \\\relax
        [\si{\kelvin}] & [\si{\mole\per\liter}] & [\si{\milli\eV\per\angstrom}] & [\si{\milli\eV} at.$^{-1}$] & [\si{\kelvin}] & [\si{\mole\per\liter}] & [\si{\milli\eV\per\angstrom}] & [\si{\milli\eV} at.$^{-1}$]\\
        \midrule
        \multicolumn{4}{l|}{\textsl{training dataset size: 200 frames}} & & & & \\
        303 & 0.56  & 38.9 & 3.5 & 343 & 0.56  & 34.4 & 2.3 \\
            & 2.32  & 40.8 & 4.4 &     & 2.32  & 34.6 & 0.2 \\
            & 6.26  & 41.3 & 1.2 &     & 6.26  & 44.6 & 1.0 \\
            & 17.89 & 35.5 & 1.8 &     & 17.89 & 39.5 & 1.4 \\
        323 & 0.56  & 37.3 & 2.1 & 363 & 0.56  & 38.0 & 2.7 \\
            & 2.32  & 36.3 & 2.9 &     & 2.32  & 44.3 & 2.1 \\
            & 6.26  & 40.4 & 2.1 &     & 6.26  & 44.4 & 5.7 \\
            & 17.89 & 39.3 & 2.0 &     & 17.89 & 38.3 & 5.3 \\
            \midrule
        \multicolumn{4}{l|}{\textsl{training dataset size: 2000 frames}} & & & & \\
        333 & 0.56  & 30.3 & 0.7 & 333 & 6.26  & 30.8 & 3.1 \\
            & 1.13  & 33.3 & 2.0 &     & 12.25 & 30.4 & 0.8 \\
            & 2.32  & 35.1 & 2.0 &     & 17.89 & 29.3 & 2.9 \\

        \bottomrule
    \end{tabular}
    }
\end{table}

To evaluate the transferability of these fine-tuned models for simulating aqueous potassium hydroxide solutions, we conducted a series of molecular dynamics simulations (details in the Supporting Information, Table \ref{tab:mace_runs}). These simulations followed the methodology described in \nameref{sec:comp_detail}. Throughout the subsequent discussion, we refer to the fine-tuned MACE-MP-0 model as MACE-FT.

We analyzed the dynamical properties of the system by calculating the mean square displacements (MSD) of different particles in the AIMD, MACE-MP-0, and MACE-FT trajectories using Equation \ref{eq:msd}. Here, $\vec{r}_{i}(t)$ represents the position of particle $i$ at time $t$. The $\mathrm{MSD}(\tau)$ is calculated by averaging over all particles $i$ and all time origins $t$ that satisfy $t+\tau<t_\mathrm{MD}$, where $t_\mathrm{MD}$ denotes the total simulation duration.

\begin{equation}
\mathrm{MSD}(\tau) = \langle \left\vert \vec{r}_{i}(t+\tau) - \vec{r}_{i}(t) \right\vert^{2} \rangle_{t,i}
\label{eq:msd}
\end{equation}

From the mean square displacements, we determined the diffusion coefficient $D$, which quantifies particle mobility, using Equation \ref{eq:diff_coeff}. To ensure reliable estimates, we calculated $D$ values by performing linear fits to the $\mathrm{MSD}(\tau)$ data within the time interval $\tau \in$ [\SI{10}{\pico\second}, \SI{30}{\pico\second}].

\begin{equation}
    D = \dfrac{1}{6} \frac{d}{d \tau}\mathrm{MSD}(\tau)
    \label{eq:diff_coeff}
\end{equation}
\FloatBarrier

A basic requirement for calculating the mean squared displacement (MSD) is knowledge of the particle positions at each time step. Positions of atomic species such as K or O are directly accessible from MD simulations. However, determining the exact position of hydroxide ions is more challenging. We distinguish between hydroxide ions and water molecules by counting the number of covalently bonded protons per oxygen atom. If one or two protons are bonded to an oxygen atom, it is classified as a hydroxide ion or a water molecule, respectively. In our approach, a covalent bond between a hydrogen and an oxygen atom is assumed if the oxygen atom is the nearest oxygen neighbor to the hydrogen atom.

The position of a hydroxide ion is identified as the position of its associated oxygen atom. During the MD simulation, the nearest oxygen neighbor of a hydrogen atom may change, leading to interconversion between hydroxide and water molecules. In such cases, the position of the hydroxide ion "jumps" from one oxygen atom to another, meaning that a hydroxide ion is not permanently associated with a single oxygen atom. This behavior aligns with the central concept of the Grotthuss mechanism.

Although more complex collective variables could be used to describe the positions of hydroxide or hydronium ions, our approach offers several advantages:
(i) it enables the definition of the positions of multiple hydroxide ions within the simulation box,
(ii) it allows the detection of diffusion pathways of individual hydroxide ions across a network of oxygen atoms, and
(iii) it facilitates the calculation of hydroxide–hydroxide radial distribution functions. The latter are presented and discussed at the end of this article, revealing surprising insights into the interactions between hydroxide ions at very high concentrations. \cite{kabbe2014}

\section*{Results}
The fine-tuned MACE-MP-0 models for aqueous potassium hydroxide solutions (concentrations ranging from 0.56 to \SI{17.89}{\mol\per\liter}) accurately reproduce the energies and forces of their respective training systems. Models trained on the larger dataset (2000 frames) demonstrate superior accuracy, with an average force error of \SI{31.6}{\milli\eV\per\angstrom} and an average energy error per atom of \SI{1.9}{\milli\eV} (see Table \ref{tab:mace_err}).

\begin{figure*}[]
    \centering
    \includegraphics[width=1\textwidth]{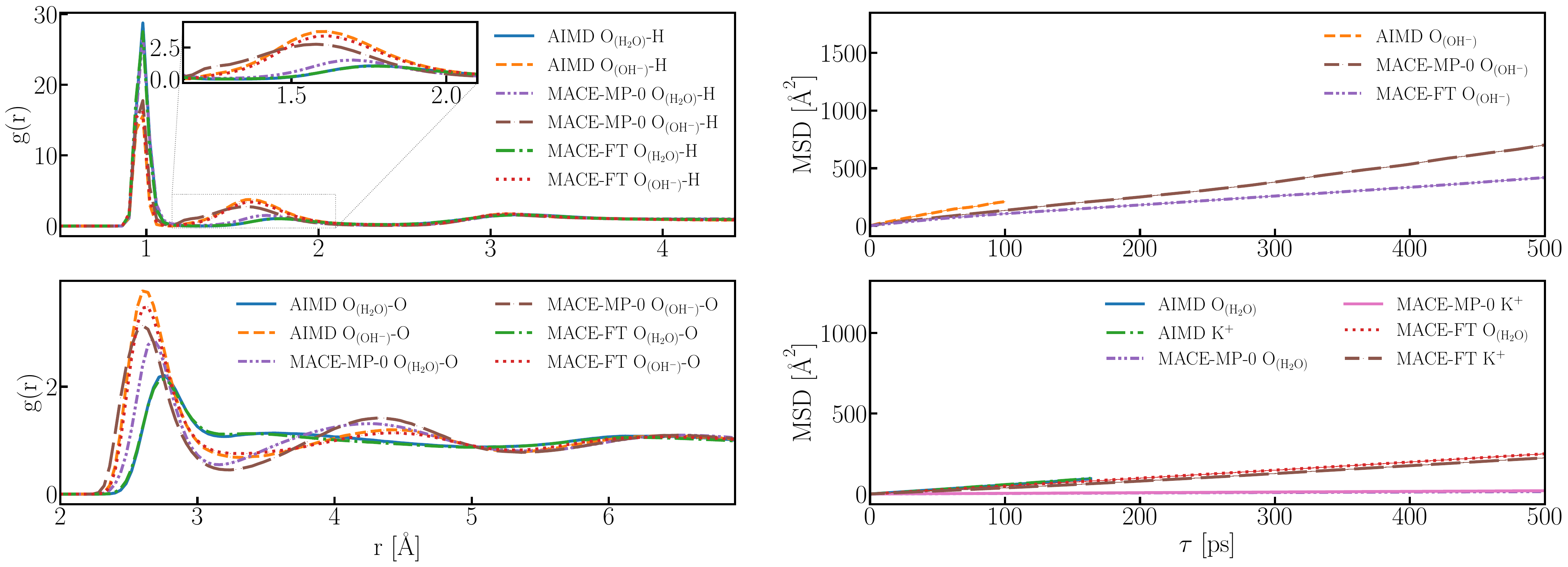}
    \caption{Comparative analysis of AIMD, MACE-MP-0, and MACE-FT trajectories with $c(\mathrm{KOH})=$ \SI{6.26}{\mole\per\liter} (RDF) and $c(\mathrm{KOH})=$ \SI{2.32}{\mole\per\liter} (MSD) at \SI{333}{\kelvin}. \textbf{Left panel}: Radial distribution functions showing \ce{O-H} (upper) and \ce{O-O} (lower) distances in water molecules and hydroxide ions. \textbf{Right panel}: Mean square displacement of hydroxide moieties (upper), potassium ions, and oxygen atoms (lower) in water molecules.}
    \label{img:rep_rdf+msd}
\end{figure*}

Molecular dynamics simulations using all of these fine-tuned models show excellent agreement with reference AIMD data. The radial distribution functions (RDFs) for \ce{O-O} and \ce{O-H} distances (Figure \ref{img:rep_rdf+msd}, left panel) confirm that the fine-tuned models accurately capture the hydrogen-bond network structure. In contrast, the non-fine-tuned foundation model (MACE-MP-0) shows significantly poorer agreement. The mean square displacement analysis for potassium ions, water oxygen atoms, and hydroxide ions (Figure \ref{img:rep_rdf+msd}, right panel) demonstrates that MD simulations with fine-tuned models can be extended to much longer timescales than computationally expensive AIMD simulations. Since AIMD simulations are inherently limited in duration, they cannot fully capture converged proton transport mechanisms, resulting in incompletely converged diffusion coefficients. This allows only qualitative comparisons with fine-tuned MD results. Notably, the non-fine-tuned foundation model misrepresents the dynamical behavior especially of \ce{K+} and the Oxygen atoms in the water molecules.

\begin{figure*}[]
    \centering
    \includegraphics[width=0.45\textwidth]{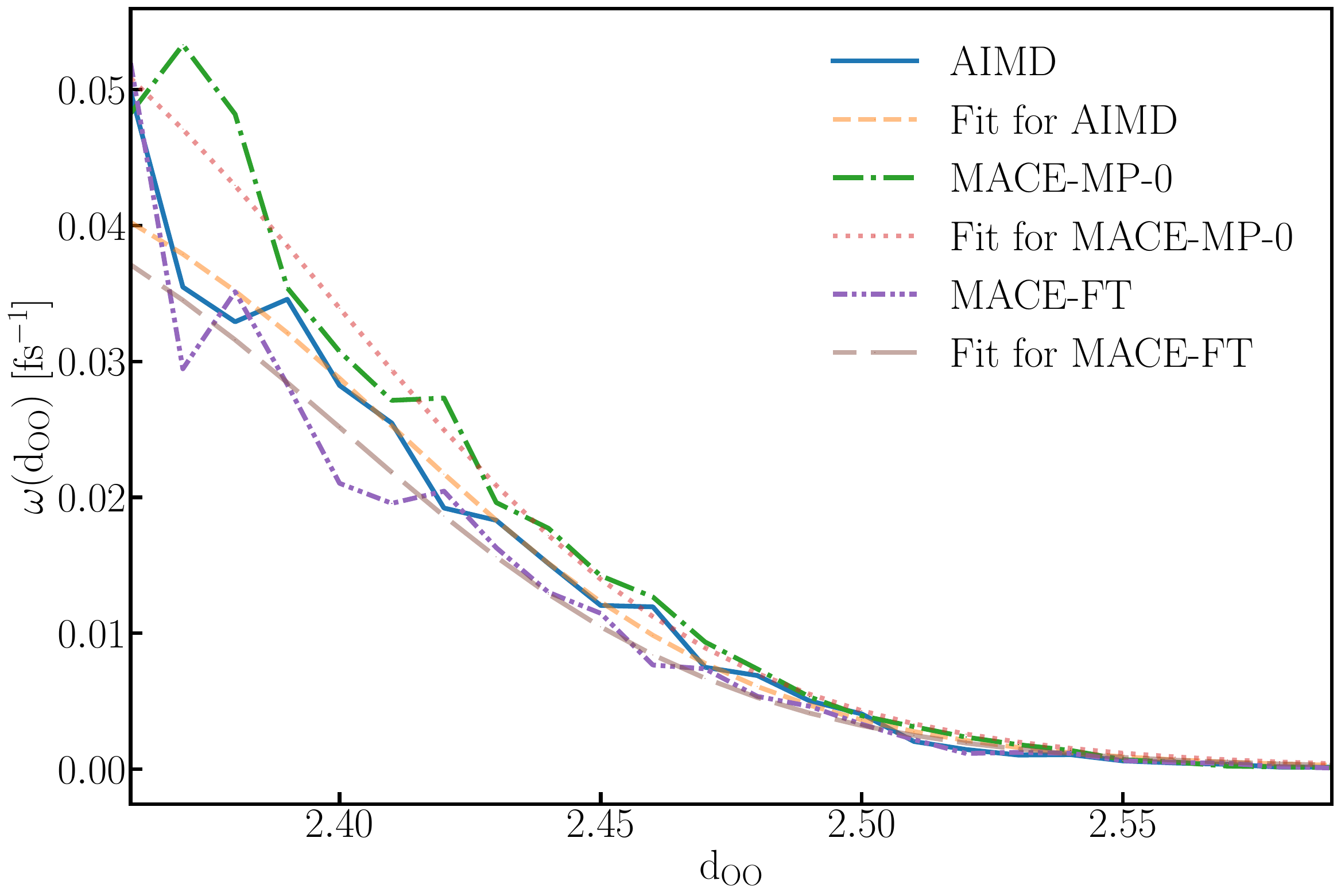}
    \hspace{0.03\textwidth}
    \includegraphics[width=0.45\textwidth]{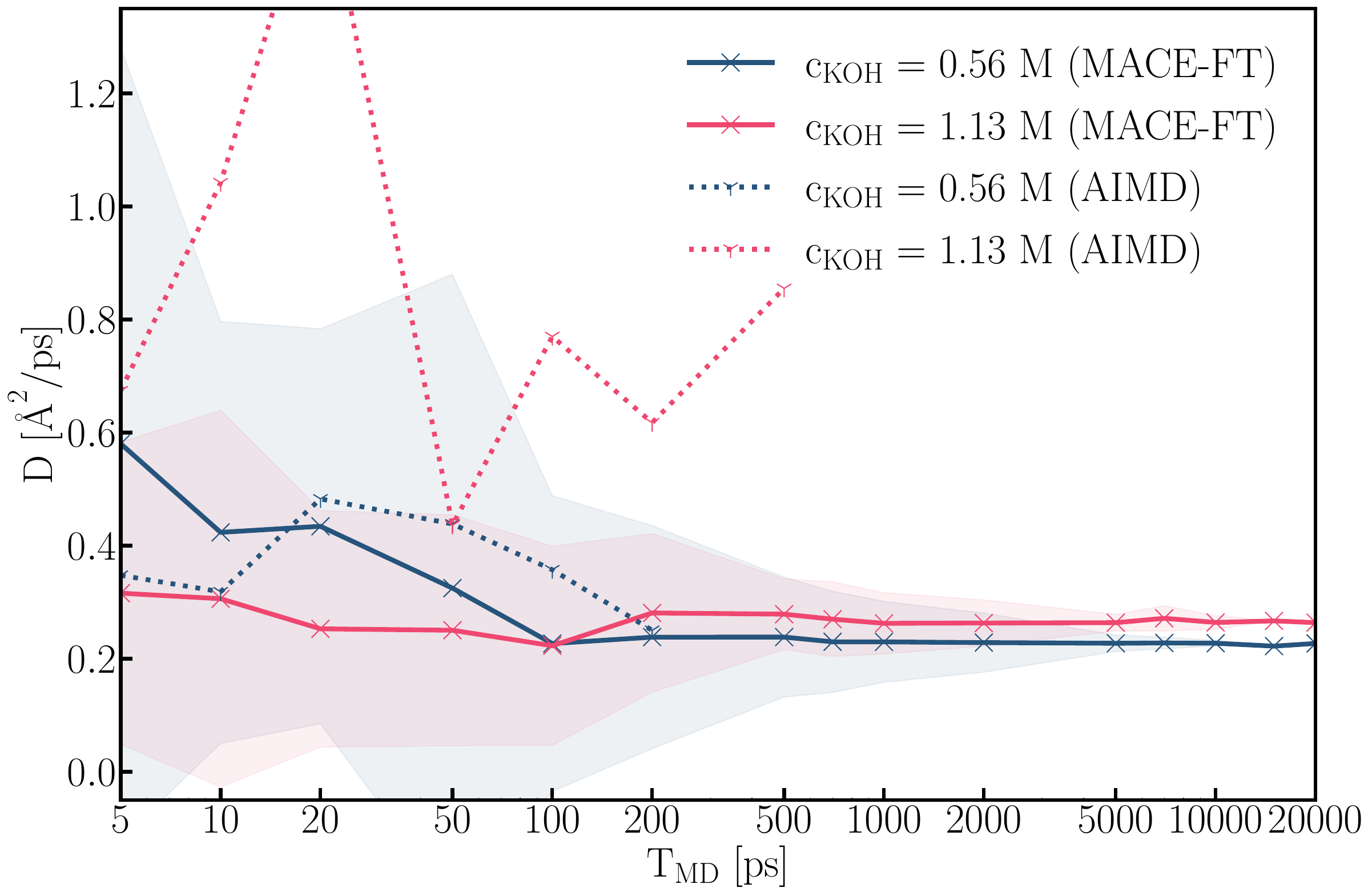}
    \caption{\textbf{Left panel}: Proton jump probability as a function of \ce{O-O} distance between hydroxide ions and neighboring water molecules. Jump rate functions were sampled from AIMD, MACE-Foundation model (MP-0), and MACE-Fine-tuned model (FT) trajectories of aqueous KOH solution with $c(\mathrm{KOH})=$ \SI{2.32}{\mole\per\liter} at \SI{333}{\kelvin}. \textbf{Right panel}: Diffusion coefficient of the \ce{OH-} moieties obtained from AIMD and MACE-FT simulations in respect to the length of the trajectory. The shaded areas represent the uncertainty of the MACE-FT D(\ce{OH-}). The presentation of the AIMD's D(\ce{OH-}) uncertainty has been omitted for reasons of clarity.}
    \label{img:rep_jr}
\end{figure*}

Previous studies have demonstrated that proton transfer within the oxygen lattice is predominantly governed by the oxygen–oxygen distance.\cite{kabbe2014, kabbe2016, kabbe2017, dressler2016, haenseroth2025ohlmc} Here, we define a proton jump as a change in the nearest oxygen neighbor between two consecutive time steps.  Based on this definition, we analyzed proton jump behavior in AIMD, MACE-MP-0, and MACE-FT trajectories using established protocols.\cite{kabbe2014, kabbe2016, kabbe2017, dressler2016, haenseroth2025ohlmc}
Our results confirm that the relationship between proton transfer probabilities and oxygen-oxygen distances is accurately described by a Fermi-like function (Figure \ref{img:rep_jr}, describable by Equation \ref{eq:fermi}).
\begin{equation} 
    \omega(\mathrm{d}_{ij}) = \frac{a}{1 + \exp\left(\frac{\mathrm{d}_{ij}-b}{c}\right)}
    \label{eq:fermi}
\end{equation}

For each simulation, we computed the conditional probability of a proton jump occurring at a specific oxygen-oxygen distance, $\omega(\mathrm{d}_{\mathrm{OO}})$, by counting actual proton jumps at that distance and dividing by the total occurrences of that \ce{O-O} distance between hydroxide ions and water molecules.

Importantly, an accurate description of the rate function $\omega(\mathrm{d}_{\mathrm{OO}})$ requires only a modest number of proton jumps per distance window within the \textit{ab initio} trajectory - a fully converged proton diffusion statistic from extended \textit{ab initio} simulations is not necessary. By comparing jump rate functions and their corresponding Fermi fit parameters (Figure \ref{img:rep_jr}, Table \ref{tab:fermi_fit}) from AIMD and MACE-FT trajectories, we gain critical insights into proton transfer dynamics. The AIMD and MACE-FT jump rate functions exhibit nearly identical proton transfer behavior, while the MACE-MP-0 jump rate function shows notable differences.

\begin{table}[!ht]
    \centering
    \caption{Fermi fit parameters describing the jump rate function of the aqueous KOH solution with $c(\mathrm{KOH})=$ \SI{2.32}{\mole\per\liter} at \SI{333}{\kelvin}.}
    \label{tab:fermi_fit}
    \begin{tabular}{l l l l}
        \toprule
        simulation/model  &    $a$ [\si{\per\femto\second}]     &   $b$ [\si{\angstrom}]   & $c$ [\si{\per\angstrom}]  \\
        \midrule
        AIMD    &   0.049 & 2.4 & 29 \\
        MACE-MP-0 & 0.069 & 2.4 & 27 \\
        MACE-FT &   0.049 & 2.4 & 27 \\
        \bottomrule
    \end{tabular}
\end{table}

Since AIMD trajectories are too short to obtain converged diffusion coefficients across different concentrations, we conducted extended molecular dynamics simulations (\SI{20}{\nano\second}) for the two most dilute systems using our fine-tuned models. The non-converged hydroxide diffusion coefficients from first-principles simulations (200-\SI{500}{\pico\second} duration) were $D$\textsubscript{0.56 M}(\ce{OH-}) = \SI{0.25}{\square\angstrom\per\pico\second} and $D$\textsubscript{1.13 M}(\ce{OH-}) = \SI{0.86}{\square\angstrom\per\pico\second}. In comparison, the values obtained from MACE-FT trajectories were $D$\textsubscript{0.56 M}(\ce{OH-}) = \SI{0.23}{\square\angstrom\per\pico\second} and $D$\textsubscript{1.13 M}(\ce{OH-}) = \SI{0.26}{\square\angstrom\per\pico\second}. While the AIMD results show substantial variation between the two concentrations, the MACE-FT models (trained on only 2000 frames from these AIMD simulations) yield more consistent diffusion coefficients.

\subsection*{Transferability across concentrations}

Thus far, our analyses have focused on MD simulations using models fine-tuned on AIMD trajectories with matching KOH concentration and temperature. We now examine the transferability of these models to different concentrations and temperatures.

To assess concentration transferability, we compared force errors against quantum chemical reference data for models trained on AIMD trajectories with $c(\mathrm{KOH})$ ranging from \SI{0.56}{\mole\per\liter} to \SI{17.89}{\mole\per\liter} at \SI{333}{\kelvin}, evaluating their performance in predicting structural behavior and diffusion coefficients at different concentrations (simulation protocol in Supporting Information Table \ref{tab:mace_runs}).

\begin{figure*}[]
    \centering
    \includegraphics[width=\textwidth]{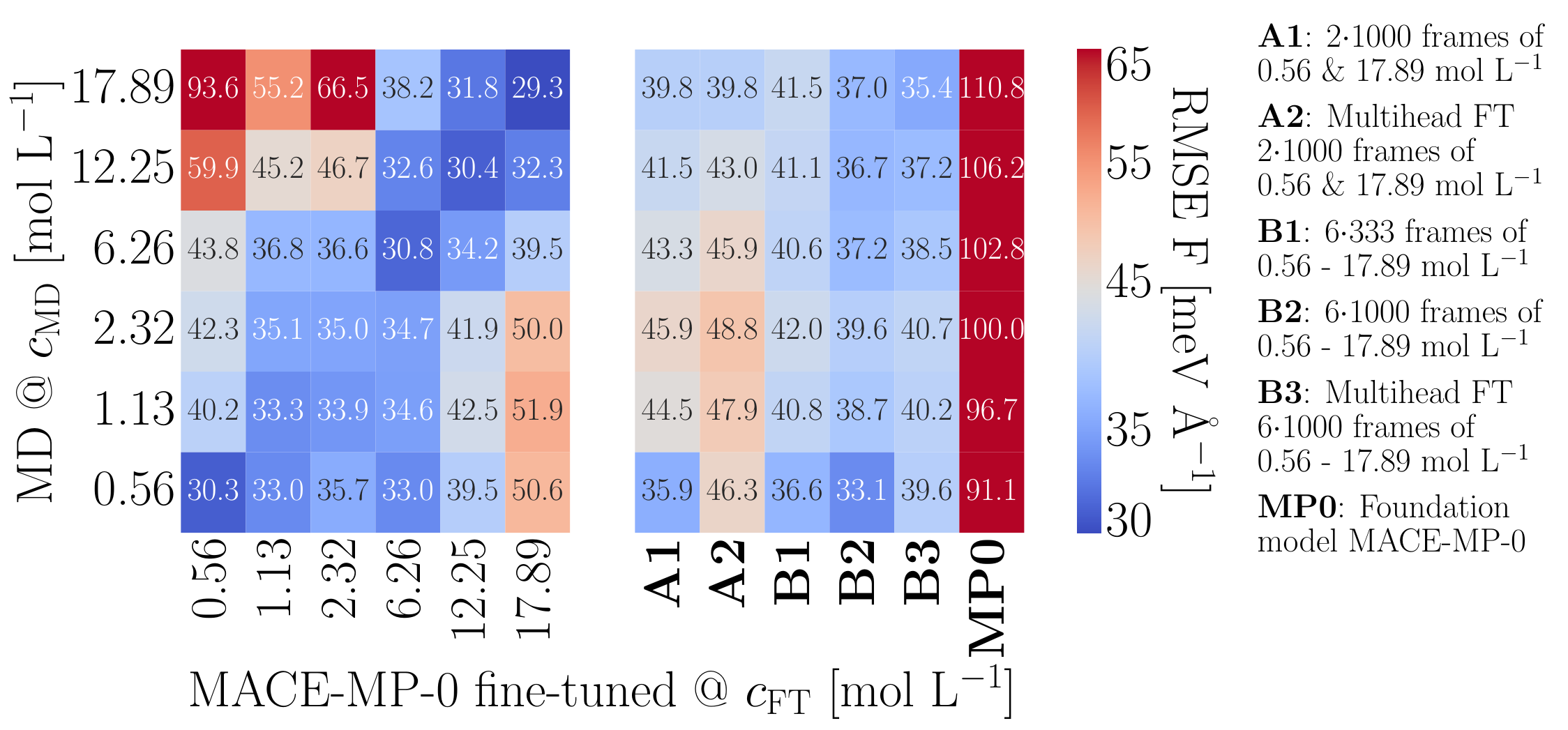}
    \caption{Force errors (RMSE, \si{\milli\eV\per\angstrom}) of fine-tuned MACE-MP-0 models across different KOH concentrations. \textbf{Left block}: Models trained on single-concentration datasets. \textbf{Right block}: Models trained on combined datasets (A1-2, B1-3). MP-0 column: Force errors from the non-fine-tuned foundation model. Color scale is capped at \SI{65}{\milli\eV\per\angstrom}, with higher values shown in the same deep red.} 
    \label{img:transfer_conc}
\end{figure*}

Figure \ref{img:transfer_conc} presents a comprehensive analysis of force prediction errors for MACE-MP-0 models fine-tuned on various KOH concentration datasets. The y-axis indicates the concentration used in the evaluation MD trajectory, while the x-axis shows the concentration used for model fine-tuning. The MP0 column (rightmost) represents the non-fine-tuned foundation model, which exhibits consistently high errors across all concentrations, underscoring the necessity of fine-tuning.

In the left block, models fine-tuned on individual concentrations reveal several key insights:
\begin{enumerate}
    \item Diagonal elements (matching training and evaluation concentrations) show almost for every concentration the lowest errors, confirming the advantage of concentration-matched training.

    \item Models trained on extreme concentrations (\SI{0.56}{\mole\per\liter} and \SI{17.89}{\mole\per\liter}) perform well at their respective training concentration but show high errors at the respective other extreme concentration.
    
    \item The fine-tuned models on extreme concentrations show asymmetric transferability patterns. The \SI{0.56}{\mole\per\liter} model performs poorly at high concentrations (error of \SI{93.6}{\milli\eV\per\angstrom} at \SI{17.89}{\mole\per\liter}), while the \SI{17.89}{\mole\per\liter} model maintains moderate accuracy at low concentrations (error of \SI{50.6}{\milli\eV\per\angstrom} at \SI{0.56}{\mole\per\liter}). This asymmetry likely stems from dataset composition differences. The \SI{17.89}{\mole\per\liter} dataset contains 25 \ce{OH-}, 25 \ce{K+}, and 50 \ce{H2O} per frame, providing rich examples of ion-ion and ion-water interactions. In contrast, the \SI{0.56}{\mole\per\liter} dataset has only 1 \ce{OH-}, 1 \ce{K+}, and 98 \ce{H2O} per frame, offering limited examples of ionic interactions.

    \item Models trained on intermediate concentrations (\SI{2.32}{\mole\per\liter}, \SI{6.26}{\mole\per\liter}, and \SI{12.25}{\mole\per\liter}) show better transferability across the concentration range. Particularly notable is the \SI{6.26}{\mole\per\liter} model, with errors ranging from \SI{30.8}{\milli\eV\per\angstrom} (at \SI{6.26}{\mole\per\liter}) to \SI{38.2}{\milli\eV\per\angstrom} (at \SI{17.89}{\mole\per\liter}) and \SI{33.0}{\milli\eV\per\angstrom} (at \SI{0.56}{\mole\per\liter}), demonstrating exceptional balance between specialization and transferability.
    
\end{enumerate}

The right block shows models (A1–B3) fine-tuned on combined datasets from multiple concentrations:
\begin{enumerate}
    \item Models A1 and A2 use data from the lowest and highest concentrations, with A2 employing multihead fine-tuning to better capture these extremes. While they demonstrate improved transferability compared to single-concentration models, their performance remains limited. The training set of these models have the same size as the single-concentration models.

    \item Model B1, fine-tuned with equal amounts of data from all six concentrations, shows balanced force errors across all concentrations, but these errors exceed those of the best single-concentration model at each evaluation point.

    \item Models B2 and B3, trained on larger combined datasets (6000 frames, compared to 2000 for single-concentration models), achieve enhanced performance across all concentrations, though at significantly higher computational cost—requiring six separate AIMD simulations and triple the fine-tuning resources.
\end{enumerate}

Remarkably, the single-concentration model trained on \SI{6.26}{\mole\per\liter} outperforms most of the more computationally intensive combined-dataset models. This suggests that a well-chosen single-concentration model can be as effective as sophisticated multi-concentration models, particularly when the target concentration range is known. For systems with highly variable concentrations, the combined-dataset models (B2 and B3) may justify their higher computational cost, but our findings challenge the assumption that training data  from  a wide range of systems is always necessary for robust transferability in complex ionic solutions.

The transferability of these models is further evaluated by examining hydroxide jump dynamics. Figure \ref{img:no_jumps_conc} shows the relative difference in jumps per picosecond per hydroxide between simulations using fine-tuned models and reference AIMD simulations at the same concentration. Models fine-tuned on dilute solutions significantly underestimate hydroxide jumps at higher concentrations (by up to \SI{45}{\percent}), while models trained on concentrated solutions overestimate jumps at lower concentrations (by about \SI{15}{\percent}). This discrepancy likely stems from the rarity of hydroxide transfer events in dilute system simulations, due to fewer hydroxide moieties in the training data. Models trained on intermediate concentrations show more balanced performance with smaller deviations from AIMD results.

To further evaluate model transferability, we investigated the ability of different models to capture hydroxide–hydroxide interactions - a phenomenon not explicitly represented in the dilute training data. Hydroxide ions generally repel one another due to their negative charges. However, they can still participate in hydrogen-bonding networks, which modulate these repulsive interactions. In aqueous environments, water molecules help screen the electrostatic repulsion between hydroxide ions. As the concentration increases, particularly in highly alkaline solutions, this screening becomes less effective, leading to stronger and more structured hydroxide–hydroxide interactions. Under extremely concentrated conditions, where water is scarce, two hydroxide ions can form a weakly stabilized pair by sharing a hydrogen bond through their hydrogen atoms and lone electron pairs. This phenomenon was first reported by Coste et al. (2019), who presented hydroxide–hydroxide radial distribution functions for aqueous sodium hydroxide solutions at concentrations up to \SI{9.7}{\mole\per\liter} (see their Figure 2d).\cite{coste2019} At the highest concentration, they observed a shoulder to the left of the first peak at \SI{3.3}{\angstrom}, which they attributed to "the formation of an H-bond between two hydroxide anions". In this work, we discuss such an hydrogen bond between two hydroxide ions in an aqueous potassium hydroxide solution for the first time. At even higher concentrations, this shoulder evolves into a distinct peak.

The hydroxide–hydroxide radial distribution functions, calculated from AIMD trajectories of aq. KOH solution at different concentrations (see Figure \ref{img:rdfoo_conc}), reveal increased hydrogen bonding between hydroxide ions at higher concentrations, evidenced by emerging peaks at $\mathrm{d}$\textsubscript{OO} $<$ \SI{3}{\angstrom}. While no indication of hydrogen bonding between hydroxide ions is observed below \SI{6.26}{\mole\per\liter}, hydrogen bonding becomes the dominant feature in the RDF at \SI{17.89}{\mole\per\liter}. The hydroxide–hydroxide RDFs calculated from the AIMD trajectories at various concentrations are provided in the Supporting Information.

We utilize their concentration-dependent formation as a sensitive, MD-relevant observable to evaluate the transferability of machine-learned force fields across different hydroxide concentrations. The radial distribution functions for the O\textsubscript{(\ce{OH-})}-O\textsubscript{(\ce{OH-})} interaction in a highly concentrated system are shown in the lower panel of Figure \ref{img:boxes}. These RDFs were obtained from trajectories of aq. KOH solution at \SI{17.89}{\mole\per\liter} generated by AIMD, the MACE-MP-0 foundation model and the MACE fine-tuned models.
The MACE-MP-0 foundation model predicts the interaction behavior at short distances (where two \ce{OH-} ions are separated by one hydrogen bond) but shows an offset toward smaller oxygen-oxygen distances compared to the AIMD data. The predicted intensity of this peak is also too large. Similarly, the RDF from the MACE model fine-tuned on the \SI{1.13}{\mole\per\liter} AIMD trajectory also exhibits the same offset by the AIMD data at short distances. Interestingly, the model trained on the most dilute system (\SI{0.56}{\mole\per\liter}) predicts hydroxide-hydroxide interactions well, even though such interactions were absent in its training data (as the training set contained only one \ce{OH-} ion per simulation box). 
Models trained on simulations of highly concentrated aqueous potassium hydroxide solutions show better agreement with the AIMD RDF. The best match with the AIMD data is achieved by the model fine-tuned on the \SI{12.25}{\mole\per\liter} trajectory. However, the model trained on the highest concentration slightly overestimates the intensity of the first peak. The minimum around \SI{3.4}{\angstrom} is well captured by all fine-tuned models, as is the subsequent peak near \SI{4.5}{\angstrom}. Beyond this point, models fine-tuned on more dilute systems again show poorer agreement with the AIMD data.

We evaluated temperature transferability across five temperatures (303–\SI{363}{\kelvin}) for four different models (0.56, 2.32, 6.26, and \SI{17.89}{\mol\per\liter}). Figure \ref{img:transfer_temp} shows four heatmaps representing force errors for each concentration.
Unlike the concentration transferability results, temperature transferability shows relatively narrow error ranges. The force errors generally increase with higher simulation temperatures, with the lowest errors typically observed at \SI{303}{\kelvin}. This trend likely reflects the increased molecular motion and more diverse configurational sampling at elevated temperatures, which presents a greater challenge for accurate force prediction. Models fine-tuned at low temperatures show higher errors when applied to high-temperature MDs, presumably because they have not encountered the broader phase space sampled at higher thermal energies. Conversely, models trained at intermediate temperatures exhibit balanced performance across the temperature range. 

\section*{Conclusion}
Our study demonstrates that fine-tuned machine-learning force fields enable simulations of hydroxide-ion transport in aqueous potassium-hydroxide solutions over a wide range of concentrations and temperatures.
The MACE-MP-0 foundation model, when fine-tuned on relatively small datasets from first-principles molecular dynamics simulations, achieves remarkable accuracy and transferability, enabling simulations at the extended timescales necessary to fully characterize hydroxide ion dynamics.

Two key findings emerge from our investigation. First, models fine-tuned on intermediate concentrations (particularly at \SI{6.26}{\mole\per\liter}) exhibit exceptional transferability across the entire concentration range, often outperforming more complex multi-concentration models while requiring significantly less computational resources.

Second, at high KOH concentrations we observe the emergence of hydroxide–hydroxide hydrogen bonding—a phenomenon previously reported only in sodium hydroxide solutions and absent from our low-concentration training data. Remarkably, MLFF models fine-tuned on those low-concentration KOH systems still predict this unusual bonding pattern when applied to high-concentration simulations.

The temperature transferability of our models further demonstrates their robustness, with relatively narrow error ranges observed across different operating temperatures. This suggests that models trained at intermediate temperatures can effectively capture hydroxide dynamics across the temperature range relevant to electrolyzer operation.

Our approach establishes a framework for developing broadly applicable machine learning force fields capable of simulating AEM materials at extended time and length scales. By bridging the gap between theoretical models and practical operating conditions, these simulations will contribute to the development of electrolyzers for green hydrogen production.

\FloatBarrier
\section*{Computational Details} \label{sec:comp_detail}
\subsection*{\textit{ab initio} Molecular Dynamics Simulation}
We simulated aqueous potassium hydroxide solutions with concentrations ranging from \SI{0.56}{\mole\per\liter} to \SI{17.89}{\mole\per\liter} at temperatures between at \SI{333}{\kelvin}. Details of these simulations are provided in Tables \ref{tab:aimd_KOH}. Additional AIMD simulations were performed on potassium hydroxide solutions across a range of concentrations ($c$(KOH) = 0.56, 2.32, 6.26, and \SI{17.89}{\mole\per\liter}) and temperatures (303, 323, 343, and \SI{363}{\kelvin}) with a simulation time of \SI{50}{\pico\second}.

\begin{table}[]
    \centering
    \caption{Computational details of the AIMD simulations at \SI{333}{\kelvin}.}
        \begin{tabular}{l l l l}
        \toprule
        simulation of   &	1 KOH              &  2 KOH            & 4 KOH         \\
        			    &	in 98 \ch{H2O}     &  in 96 \ch{H2O}   & in 92 \ch{H2O} \\
        \midrule
        $c$(KOH) [\si{\mole\per\liter}]   & 0.56 & 1.13 & 2.32  \\
        w(KOH) [\si{\percent}]          & 3 & 6 & 12 \\ 
        cubic box vector [\si{\angstrom}]   & 14.41  & 14.34 & 14.21   \\
        angle [\si{\degree}]    & $\alpha=90$ & $\beta=90$ & $\gamma=90$ \\
        number of atoms	        & 297& 294 & 270 \\
        time step [\si{\femto\second}]  	& 0.5 & 0.5 & 0.5 \\
        temperature	[\si{\kelvin}]	& 333  & 333 & 333 \\
        simulation time [\si{\pico\second}]	& 300 & 500 & 250  \\
        \bottomrule					
               \toprule
        simulation of   &	10 KOH              &	18 KOH 	          &  25 KOH                            \\
        			    &	in 80 \ch{H2O}     &    in 64 \ch{H2O} 	  &  in 50 \ch{H2O}        \\
        \midrule
        $c$(KOH) [\si{\mole\per\liter}]   & 6.26 & 12.25 & 17.89   \\
        w(KOH) [\si{\percent}]          & 28 & 48 & 61    \\ 
        cubic box vector [\si{\angstrom}] & 13.85 & 13.46 & 13.24    \\
        angle [\si{\degree}]    & $\alpha=90$ & $\beta=90$ & $\gamma=90$  \\
        number of atoms	        & 270 & 246 & 225   \\
        time step [\si{\femto\second}]  	& 0.5 & 0.5 & 0.5  \\
        temperature	[\si{\kelvin}]	& 333  & 333 & 333   \\
        simulation time [\si{\pico\second}]	& 250 & 300 & 400 \\
        \bottomrule		
    \end{tabular}
    \label{tab:aimd_KOH}
\end{table}

Before conducting the simulations, all structures underwent geometry optimization. The \textit{ab initio} molecular dynamics (AIMD) simulations for several concentrations were performed using the CP2K software package,\cite{cp2k_1, cp2k_2, cp2k_3, kuhne2020cp2k} with a simulation temperature of \SI{333}{\kelvin}. The trajectories extended beyond \SI{250}{\pico\second}, employing a time step of \SI{0.5}{\femto\second}.

The electronic structure calculations were conducted with the density functional theory (DFT),\cite{DFT1-Theorem_Hohenberg, DFT2_Kohn_Sham1, DFT3_Kohn_Sham2} using the Quickstep module\cite{cp2k_quickstep} in CP2K. For efficient orbital transformations and fast convergence, we applied the BLYP exchange-correlation functional,\cite{cp2k_blyp1, cp2k_blyp2} the DZVP-MOLOPT-GTH basis set,\cite{cp2k_basis-set} and GTH-BLYP pseudopotentials.\cite{cp2k_gth-pseudopot1, cp2k_gth-pseudopot2, cp2k_gth-pseudopot3} Additionally, empirical dispersion corrections following the Grimme D3 method\cite{cp2k_d3, cp2k_d3_2} were included. The temperature was maintained using a Nosé-Hoover Chain thermostat within an NVT ensemble.\cite{nose1970, nose1984, martyna1992} 

\subsection*{Fine-tune MACE-MP-0}
The fine-tuning of the MACE-MP-0 foundation model was carried out using training data extracted from the AIMD trajectories of aqueous KOH solutions (see Tables \ref{tab:aimd_KOH}). The fine-tuning procedure followed the workflow described above (see Methods section) and was executed using the MACE Python package (v0.3.6).\cite{batatia2023, jain2013, ong2015} The specific settings are summarized in Table \ref{tab:mace_ft}. Atomic reference energies $E_0$ were obtained from CP2K single-point calculations. For aqueous potassium hydroxide solutions, the first set of fine-tuned model was trained using a dataset consisting of every 200th frame from a \SI{200}{\pico\second} AIMD trajectory, resulting in 2000 frames. The second set of fine-tuned models is trained on every 200th frame from a \SI{20}{\pico\second} AIMD trajectory, resulting in 200 frames.

\begin{table}[]
    \centering
    \caption{Computational details of the fine-tuning process of MACE-MP-0.}
        \resizebox{0.48\textwidth}{!}{
        \begin{tabular}{l l l l l l l l l}
        \toprule
        foundation  & force   & energy   & training  & validation    & batch & number of & learning  \\ 
        model       & \multicolumn{2}{c}{error weight}        & dataset   & fraction      & size  & epochs    & rate \\
        \midrule
           &    &     & 200 /                           &     &   &     &        \\
        small       & 10 & 0.1 & 2000 frames  & 0.1 & 5 & 200 & 0.01   \\
                    &    &     & of AIMD                       &     &   &     &        \\
        \bottomrule		
        \end{tabular}
        }
    \label{tab:mace_ft}
\end{table}

\subsection*{Mace Molecular Dynamics Simulation}
Molecular dynamics simulations using the fine-tuned MACE-MP-0 models were performed with the ASE calculator.\cite{ase-paper} The simulation system were identical to those listed in Tables \ref{tab:aimd_KOH}. These MD simulations were run with a time step of \SI{0.5}{\femto\second} for \SI{3}{\nano\second}, maintaining constant temperatures of \SI{303}{\kelvin}, \SI{323}{\kelvin}, \SI{333}{\kelvin}, \SI{343}{\kelvin} or \SI{363}{\kelvin} using a Langevin thermostat.

\section*{Acknowledgement}

We gratefully acknowledge financial support by TAB research group "KapMemLyse".
We also thank the staff of the Compute Center of the Technische Universität Ilmenau and especially Mr.~Henning~Schwanbeck for providing an excellent research environment.

\section*{Conflict of Interest}

The authors declare no conflict of interest.

\section*{Supporting Information}

The Supporting Information is available free of charge.
\FloatBarrier

\bibliography{bibliography.bib}

\renewcommand{\thetable}{S\arabic{table}}
\renewcommand{\thefigure}{S\arabic{figure}}
\renewcommand{\thepage}{S-\Roman{page}}

\widetext
\clearpage
\FloatBarrier
\setcounter{table}{0}
\setcounter{figure}{0}
\setcounter{page}{1}
\section*{Supporting Information}

\begin{table*}[ht!]
    \centering
    \caption{Simulation protocol for all molecular dynamics simulations with the fine-tuned MACE-MP-0 models of the aqueous potassium hydroxide solutions.}
    \label{tab:mace_runs}
    \begin{tabular}{l l l l}
        \toprule
        \multicolumn{2}{c}{MACE-FT} & \multicolumn{2}{c}{MD simulation with MACE-FT of $c_\mathrm{MD}$ at T\textsubscript{MD}} \\
        T\textsubscript{FT} & $c_\mathrm{FT}$(KOH) & $c_\mathrm{MD}$ [\si{\mole\per\liter}] & $c_\mathrm{MD}$ [\si{\mole\per\liter}] \\\relax
        [\si{\kelvin}] & [\si{\mole\per\liter}] & @ T\textsubscript{MD} = T\textsubscript{FT} & @ T\textsubscript{MD} = \SI{333}{\kelvin} \\
        \midrule
        303 & 0.56      &   0.56 & 0.56 \\
            & 2.32      &   2.32 & 2.32 \\
            & 6.26      &   6.26 & 6.26 \\
            & 17.89     &   17.89 & 17.89 \\
        323 & 0.56      &   0.56 & 0.56 \\
            & 2.32      &   2.32 & 2.32 \\
            & 6.26      &   6.26 & 6.26 \\
            & 17.89     &   17.89 & 17.89 \\
        333 & 0.56      &   0.56 & 0.56, 1.13, 2.32, 6.26, 12.25, 17.89 \\
            & 1.13      &   - & 0.56, 1.13, 2.32, 6.26, 12.25, 17.89 \\
            & 2.32      &   2.32 & 0.56, 1.13, 2.32, 6.26, 12.25, 17.89 \\
            & 6.26      &   6.26 & 0.56, 1.13, 2.32, 6.26, 12.25, 17.89 \\
            & 12.25     &   - & 0.56, 1.13, 2.32, 6.26, 12.25, 17.89 \\
            & 17.89     &   17.89 & 0.56, 1.13, 2.32, 6.26, 12.25, 17.89 \\
        343 & 0.56      &   0.56 & 0.56 \\
            & 2.32      &   2.32 & 2.32 \\
            & 6.26      &   6.26 & 6.26 \\
            & 17.89     &   17.89 & 17.89 \\
        363 & 0.56      &   0.56 & 0.56 \\
            & 2.32      &   2.32 & 2.32 \\
            & 6.26      &   6.26 & 6.26 \\
            & 17.89     &   17.89 & 17.89 \\
        \bottomrule
    \end{tabular}
\end{table*}

\FloatBarrier
\begin{figure*}[]
    \centering
    \includegraphics[width=0.5\textwidth]{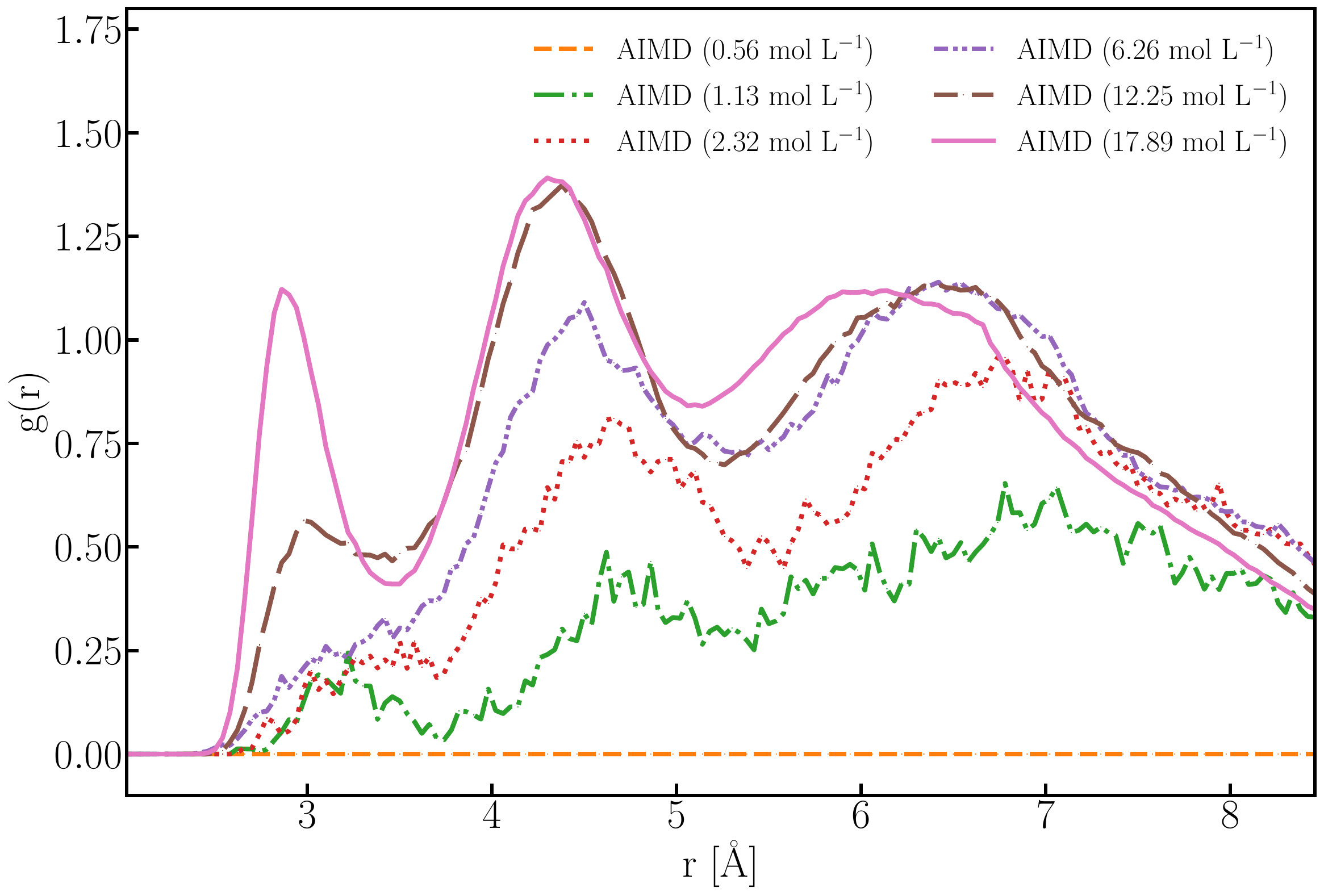}
    \caption{
    Radial distribution function of O\textsubscript{(\ce{OH-})}-O\textsubscript{(\ce{OH-})} distances in the systems with $c$(KOH)= 0.56 to \SI{17.89}{\mole\per\liter}, obtained from AIMD simulations.}
    \label{img:rdfoo_conc}
\end{figure*}

\begin{figure*}[]
    \centering
    \includegraphics[width=0.45\textwidth]{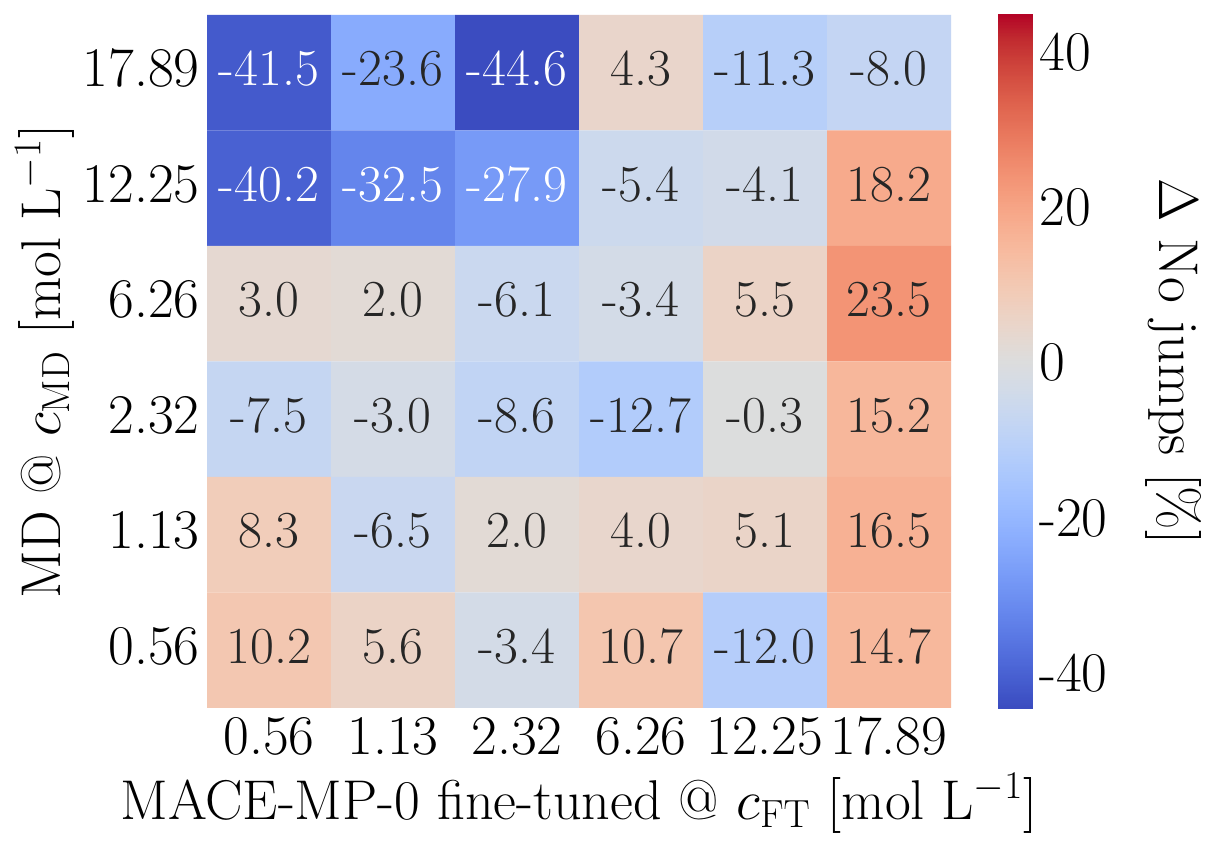}
    \caption{Relative difference in hydroxide jump frequency between fine-tuned model simulations and reference AIMD simulations across different KOH concentrations. Each model was trained on a single concentration AIMD trajectory. The heatmap reveals systematic under/overestimation patterns related to training concentration.} 
    \label{img:no_jumps_conc}
\end{figure*}

\begin{figure*}[]
    \centering
    \includegraphics[width=\textwidth]{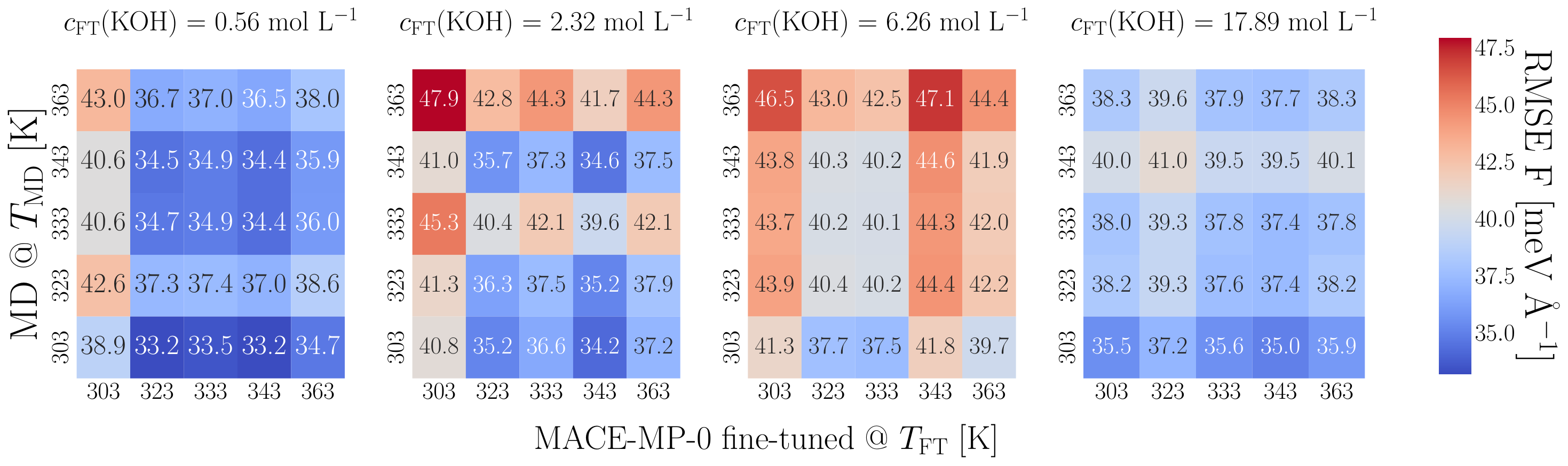}
    \caption{Force errors of fine-tuned MACE-MP-0 models (dataset size: 200 frames) evaluated at different temperatures for four KOH concentrations. Each model was trained on data at temperature T\textsubscript{FT} (x-axis) and evaluated at temperature T\textsubscript{MD} (y-axis). Force errors show less variation across temperatures than across concentrations.} 
    \label{img:transfer_temp}
\end{figure*}

\FloatBarrier

\begin{sidewaystable}[!ht]
    \caption{Root mean square error (RMSE) of force and energy of the fine-tuned MACE-MP-0\cite{batatia2023} models for aqueous potassium hydroxide solution simulating other concentrations at \SI{333}{\kelvin}.}
    \label{tab:transfer_conc}
    \resizebox{\textwidth}{!}{%
    \tiny
    \begin{tabular}{l l l l l l l l l l l l l l}
        \toprule
        MACE-FT of  &   \multicolumn{13}{c}{MD simulation with MACE-FT of $c_\mathrm{MD}$}  \\
        
        $c_\mathrm{FT}$(KOH)  & $c_\mathrm{MD}$ = & \multicolumn{2}{c}{\SI{0.56}{\mole\per\liter}} & \multicolumn{2}{c}{\SI{1.13}{\mole\per\liter}} & \multicolumn{2}{c}{\SI{2.32}{\mole\per\liter}} & \multicolumn{2}{c}{\SI{6.26}{\mole\per\liter}} & \multicolumn{2}{c}{\SI{12.25}{\mole\per\liter}} & \multicolumn{2}{c}{\SI{17.89}{\mole\per\liter}} \\\relax 
        
        [\si{\mole\per\liter}] & &  RMSE F & RMSE E & RMSE F & RMSE E & RMSE F & RMSE E & RMSE F & RMSE E & RMSE F & RMSE E & RMSE F & RMSE E \\
        & & [\si{\milli\eV\per\angstrom}] & [\si{\milli\eV} at.$^{-1}$] & [\si{\milli\eV\per\angstrom}] & [\si{\milli\eV} at.$^{-1}$] & [\si{\milli\eV\per\angstrom}] & [\si{\milli\eV} at.$^{-1}$] & [\si{\milli\eV\per\angstrom}] & [\si{\milli\eV} at.$^{-1}$] & [\si{\milli\eV\per\angstrom}] & [\si{\milli\eV} at.$^{-1}$] & [\si{\milli\eV\per\angstrom}] & [\si{\milli\eV} at.$^{-1}$] \\
        \midrule
 0.56&& 30.3 & 0.7 & 40.2 & 1430.9 & 42.3 & 4292.1 & 43.8 & 13618.6 & 59.9 & 28177.8 & 93.6 & 43499.3 \\
 1.13&& 33.0 & 454.9 & 33.3 & 1.9 & 35.1 & 852.0 & 36.8 & 3612.7 & 45.2 & 7925.5 & 55.2 & 12491.2 \\ 
 2.32&& 35.7 & 1142.4 & 33.9 & 739.3 & 35.0 & 35.1 & 36.6 & 2409.9 & 46.7 & 6170.0 & 66.5 & 10155.3 \\
 6.26&& 33.0 & 1717.2 & 34.6 & 1501.1 & 34.7 & 1144.0 & 30.8 & 3.0 & 32.6 & 1798.0 & 38.2 & 3722.1 \\
12.25&& 39.5 & 2589.0 & 42.5 & 2418.5 & 41.9 & 2155.5 & 34.2 & 1315.8 & 30.4 & 0.8 & 31.8 & 1424.5 \\
17.89&& 50.6 & 2784.8 & 51.9 & 2653.5 & 50.0 & 2471.4 & 39.5 & 1895.2 & 32.3 & 979.6 & 29.3 & 2.9 \\
Foundation Model MP-0 & & 91.1 & 153721.2 & 96.7 & 156269.4 & 100.0 & 161643.5 & 102.8 & 179090.0 & 106.2 & 206364.17 & 110.8 & 234956.9 \\
        \bottomrule
    \end{tabular}}

    \bigskip\bigskip  
    \caption{Root mean square error (RMSE) of force and energy of the fine-tuned MACE-MP-0 models for aqueous potassium hydroxide solution simulating other temperatures: \SI{303}{\kelvin}, \SI{323}{\kelvin}, \SI{333}{\kelvin}, \SI{343}{\kelvin} and \SI{363}{\kelvin}.}
    \label{tab:transfer_temp}
    \smallskip
    \resizebox{\textwidth}{!}{%
    \tiny
    \begin{tabular}{l l l l l l l l l l l l}
        \toprule
        MACE-FT of  &   &  \multicolumn{10}{c}{MD simulation with MACE-FT of $c_\mathrm{MD}\mathrm{(KOH)} = c_\mathrm{FT}\mathrm{(KOH)}$} \\
        
        $c_\mathrm{FT}$(KOH) & & \multicolumn{2}{c}{at \SI{303}{\kelvin}} & \multicolumn{2}{c}{at \SI{323}{\kelvin}} & \multicolumn{2}{c}{at \SI{333}{\kelvin}} & \multicolumn{2}{c}{at \SI{343}{\kelvin}} & \multicolumn{2}{c}{at \SI{363}{\kelvin}} \\
        
        [\si{\mole\per\liter}] & &
        RMSE E & RMSE F & RMSE E & RMSE F & RMSE E & RMSE F & RMSE E & RMSE F & RMSE E & RMSE F\\
        at T\textsubscript{FT} & & 
        [\si{\milli\eV\per\angstrom}] & [\si{\milli\eV} at.$^{-1}$] & 
        [\si{\milli\eV\per\angstrom}] & [\si{\milli\eV} at.$^{-1}$] & 
        [\si{\milli\eV\per\angstrom}] & [\si{\milli\eV} at.$^{-1}$] & 
        [\si{\milli\eV\per\angstrom}] & [\si{\milli\eV} at.$^{-1}$] & 
        [\si{\milli\eV\per\angstrom}] & [\si{\milli\eV} at.$^{-1}$] \\
        \midrule

 \multicolumn{12}{l}{T\textsubscript{FT} = \SI{303}{\kelvin}} \\
  0.56&&  38.3 & 3.5 & 42.6 & 5.0 & 40.6 & 3.8 & 40.6 & 3.8 & 43.0 & 3.5 \\
  2.32&&  40.8 & 4.4 & 41.3 & 4.0 & 45.3 & 5.4 & 41.0 & 3.6 & 47.9 & 5.7 \\  
  6.26&&  41.3 & 1.2 & 43.9 & 1.9 & 43.7 & 1.7 & 43.8 & 1.7 & 46.5 & 2.2 \\
 17.89&&  35.5 & 1.8 & 38.2 & 2.6 & 38.0 & 2.4 & 40.0 & 2.7 & 38.3 & 1.9\\

 \midrule
 \multicolumn{12}{l}{T\textsubscript{FT} = \SI{323}{\kelvin}} \\
  0.56&&  33.2 & 0.6 & 37.3 & 2.1 & 34.7 & 0.7 & 34.5 & 0.7 & 36.7 & 0.6 \\
  2.32&&  35.2 & 2.9 & 36.3 & 2.9 & 40.4 & 4.3 & 35.7 & 2.5 & 42.8 & 4.4 \\
  6.26&&  37.7 & 1.5 & 40.4 & 2.1 & 40.2 & 1.8 & 40.3 & 1.8 & 43.0 & 2.2 \\
 17.89&&  33.2 & 1.5 & 37.3 & 2.0 & 34.7 & 1.9 & 34.5 & 2.1 & 36.7 & 1.5 \\

\midrule 
 \multicolumn{12}{l}{T\textsubscript{FT} = \SI{333}{\kelvin}} \\
  0.56&& 33.5 & 1.4 & 37.4 & 2.8 & 34.9 & 1.4 & 34.9 & 1.3 & 37.0 & 1.6 \\   
  2.32&& 36.6 & 2.6 & 37.5 & 2.5 & 42.1 & 3.8 & 37.3 & 2.0 & 44.3 & 3.9 \\
  6.26&& 37.5 & 1.9 & 40.2 & 2.6 & 40.1 & 2.2 & 40.2 & 2.2 & 42.5 & 2.6 \\
 17.89&& 35.6 & 24.1& 37.6 & 2.9 & 37.8 & 2.8 & 39.5 & 3.0 & 37.9 & 2.4 \\

\midrule
 \multicolumn{12}{l}{T\textsubscript{FT} = \SI{343}{\kelvin}} \\
  0.56&& 33.2 & 2.2 & 37.0 & 3.6 & 34.4 & 2.3 & 34.4 & 2.3 & 36.5 & 2.0 \\
  2.32&& 34.2 & 0.6 & 35.2 & 0.6 & 39.6 & 1.9 & 34.6 & 0.2 & 41.7 & 2.1 \\   
  6.26&& 41.8 & 0.5 & 44.4 & 1.3 & 44.3 & 1.0 & 44.6 & 1.0 & 47.1 & 1.5 \\
 17.89&& 35.0 & 0.7 & 37.4 & 1.3 & 37.4 & 1.2 & 39.5 & 1.4 & 37.7 & 0.8 \\

 \midrule
 \multicolumn{12}{l}{T\textsubscript{FT} = \SI{363}{\kelvin}} \\
  0.56&& 34.7 & 2.6 & 38.6 & 4.0 & 36.0 & 2.6 & 35.9 & 2.4 & 38.0 & 2.7 \\
  2.32&& 37.2 & 0.5 & 37.9 & 0.7 & 42.1 & 1.9 & 37.5 & 0.3 & 44.3 & 2.1 \\
  6.26&& 39.7 & 4.6 & 42.2 & 5.4 & 42.0 & 5.1 & 41.9 & 5.1 & 44.4 & 5.7 \\
 17.89&& 35.9 & 5.2 & 38.2 & 5.7 & 37.8 & 5.6  & 40.1 & 5.9 & 38.3 & 5.3 \\
        \bottomrule
    \end{tabular}}
    
\end{sidewaystable}

\end{document}